\documentclass[10pt,a4paper]{book}
\usepackage{udineHyderabad-Klu2}

\begin{document}

	\pagestyle{fancy}
	
\talktitle{Simulating the High Energy Gamma-ray sky seen by the GLAST Large Area Telescope}{Simulating the High Energy Gamma-ray sky seen by the GLAST Large Area Telescope}

\talkauthors{F.~Longo\structure{a}, 
	P.~Azzi\structure{b}, 
	D.~Bastieri\structure{b},
	G.~Busetto\structure{b},
	Y.~Lei\structure{b},
	R.~Rando\structure{b},
	O.~Tibolla\structure{b},
	L.~Baldini\structure{c},
	M.~Kuss\structure{c},
	L.~Latronico\structure{c},
	N.~Omodei\structure{c},
	M.~Razzano\structure{c}, 
	G.~Spandre\structure{c},
	P.~Boinee\structure{d},
	A.~De~Angelis\structure{d}, 
	M.~Frailis\structure{d},
	M.~Brigida\structure{e},
	F.~Gargano\structure{e}, 
	N.~Giglietto\structure{e},
	F.~Loparco\structure{e},
	M.N.~Mazziotta\structure{e},
	C.~Cecchi\structure{f},
	P.~Lubrano\structure{f},
	F.~Marcucci\structure{f},
	M.~Pepe\structure{f},
	G.~Tosti\structure{f},
	A.~Lionetto\structure{g},
	A.~Morselli\structure{g},
	C.~Pittori\structure{g}}

\authorstucture[a]{University of Trieste and INFN, Sezione di Trieste} 
\authorstucture[b]{University of Padova and INFN, Sezione di Padova} 
\authorstucture[c]{University of Pisa and INFN, Sezione di Pisa} 
\authorstucture[e]{University of Bari and INFN, Sezione di Bari}
\authorstucture[d]{University of Udine and INFN, Sezione di Trieste}
\authorstucture[f]{University of Perugia and INFN, Sezione di Perugia}
\authorstucture[g]{University of Roma ``Tor Vergata'' and INFN, Sezione di Roma2}

\shorttitle{Simulating the GLAST sky} 

\firstauthor{F.Longo et al.}

		\index{Longo@\textsc{Longo}, F.}
		\index{Azzi@\textsc{Azzi}, P.}
		\index{Bastieri@\textsc{Bastieri}, D.}
		\index{Busetto@\textsc{Busetto}, G.}
		\index{Lei@\textsc{Lei}, Y.}
		\index{Rando@\textsc{Rando}, R.}
		\index{Tibolla@\textsc{Tibolla}, O.}
		\index{Baldini@\textsc{Baldini}, L.}
		\index{Kuss@\textsc{Kuss}, M.}
		\index{Latronico@\textsc{Latronico}, L.}
		\index{Omodei@\textsc{Omodei}, N.}
		\index{Razzano@\textsc{Razzano}, M.}
		\index{Spandre@\textsc{Spandre}, G.}
		\index{Boinee@\textsc{Boinee}, P.}
		\index{De Angelis@\textsc{De Angelis}, A.}
		\index{Frailis@\textsc{Frailis}, M.}
		\index{Brigida@\textsc{Brigida}, M.}
		\index{Gargano@\textsc{Gargano}, F.}
		\index{Giglietto@\textsc{Giglietto}, N.}
		\index{Loparco@\textsc{Loparco}, F.}
		\index{Mazziotta@\textsc{Mazziotta}, M.N.}
		\index{Cecchi@\textsc{Cecchi}, C.}
		\index{Lubrano@\textsc{Lubrano}, P.}
		\index{Marcucci@\textsc{Marcucci}, F.}
		\index{Pepe@\textsc{Pepe}, M.}
		\index{Tosti@\textsc{Tosti}, G.}
		\index{Lionetto@\textsc{Lionetto}, A.}
		\index{Morselli@\textsc{Morselli}, A.}
		\index{Pittori@\textsc{Pittori}, C.}

\begin{abstract}

This paper presents the simulation of the GLAST high energy gamma-ray telescope. 
The simulation package, written in C++, is based on the Geant4 toolkit, and it is integrated into a general framework used to process events. A detailed simulation of the electronic signals inside Silicon detectors has been provided and it is used for the particle tracking, which is handled by a dedicated software. 
A unique repository for the geometrical description of the detector has been realized using the XML language and a C++ library to access this information has been designed and implemented. A new event display based on the HepRep protocol was implemented. The full simulation was used to simulate a full week of GLAST high energy gamma-ray observations.  
This paper outlines the contribution developed by the Italian GLAST software group. 
\end{abstract}

\section{Introduction}

The Gamma-ray Large Area Space Telescope (GLAST) is an international mission that will
study the high-energy phenomena in gamma-rays universe \protect\cite{glast---}.
GLAST is scheduled for launch in mid 2007.
 
GLAST is instrumented with a hodoscope of Silicon planes with slabs of converter, followed by a calorimeter; the hodoscope is surrounded by an anticoincidence (ACD). This instrument, called the Large Area Telescope LAT, is sensitive to gamma rays in the
energy range between 30 MeV and 300 GeV.
The energy range, the field of view and the angular resolution of the GLAST LAT are vastly
improved in comparison with those of its predecessor EGRET (operating in 1991-2000), so that the LAT will provide a factor of 30 or more advance in sensitivity. 
This improvement should enable the detection of several thousands of new high-energy sources and allow the study of gamma-ray bursts and other transients, the resolution of the gamma-ray sky and diffuse emission, the search for evidence of dark matter and the detection of AGNs, pulsars and SNRs. A detailed description of the scientific goals of GLAST mission and an introduction to the experiment can be found in \protect\cite{gsd}.

\begin{figure}[!ht]
\centering
\includegraphics[width=3.in]{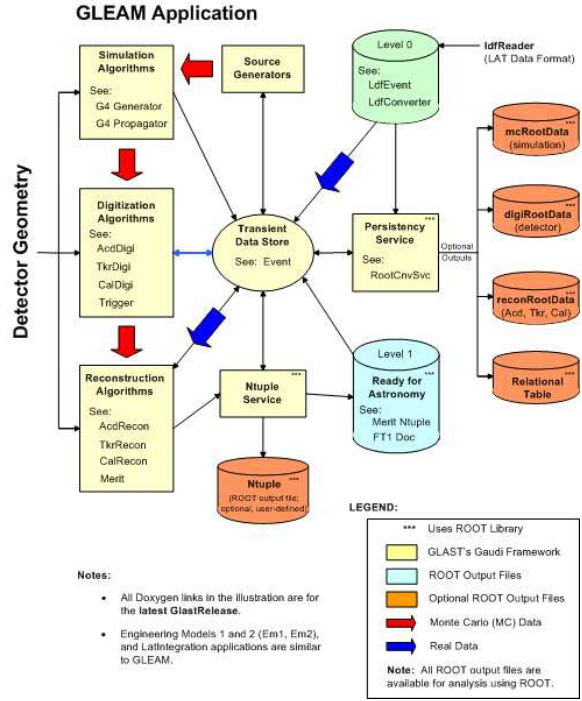}
\caption{General scheme for simulation and reconstruction within the GLAST off-line software framework}
\label{glast-sw}
\end{figure}

GLAST is a complex system, and detailed computer simulations are required 
to design the instrument, to construct the response function and to predict the background in the
orbit. To accomplish these tasks an object-oriented C++ application called
{\it Gleam} (GLAST LAT Event Analysis Machine) was adopted and implemented by the GLAST LAT collaboration. A brief description of Gleam could be found in \protect\cite{proc_pg}. Its structure is described in figure~\ref{glast-sw}. An entire week of the gamma-ray observations by GLAST-LAT was simulated using Gleam in order to develop and test GLAST LAT scientific analysis softaware.

\section{Simulation and Reconstruction Software}

The GLAST off-line software is based manly on Gaudi, a C++ framework, originally developed at CERN~\protect\cite{Gaudi}. Gaudi manages the loop of particles to be simulated, then a series of algorithms are applied to each of them to get the result of the complete simulation and reconstruction chain.
The Source Generation is the first algorithm called within the particle loop. Its task is to generate particles according to certain characteristcs. This algorithm must store the information on the temporal and spectral behaviour of the source, as well as on the orbital characteristics of GLAST. It provides a user interface to produce additional incoming particles and is responsible for setting the current time, the particle energy, direction, and type. Within this package a series of default sources are implemented. They include source for testing purposes as well as the description of astrophysical spectra and the expected particle and albedo gamma backgrounds.  An extension of this framework has been implemented for simulating transient sources such as Gamma-Ray Bursts (GRB). It can be used for studying the capability of GLAST for the observation of rapid transient fluxes in general\protect\cite{nicolaphd}. 

The algorithm which is responsible for generating the interactions of particles with the detector is based on the Geant4 MonteCarlo toolkit~\protect\cite{geant4} which is an Object Oriented (OO) simulator of the passage of particles through matter. Its application areas include high energy physics and nuclear experiments, medical science, accelerator and space physics. Within the Gleam application the simulation is managed by the  Gaudi algorithm G4Generator. 

\begin{figure}[ht!]
\centering
\includegraphics[width=3.in]{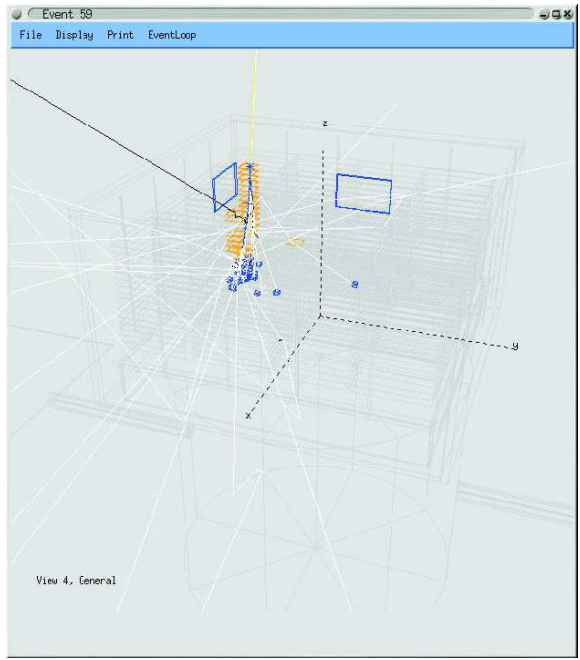}
\caption{High energy gamma-ray interacting with the GLAST LAT detector}
\label{fig--gleam}
\end{figure}

Figure~\ref{fig--gleam} shows an event generated using Geant4 within the GLAST LAT experiment. 

The next algorithm to be applied is the digitization which transforms the hits generated by the Monte Carlo simulations into the signal as read by the electronics.  
To implement a detailed digitization of the Tracker system a full simulation code has been developed\protect\cite{bari}. Starting from Monte Carlo hits in the Silicon detectors, the current signals induced on each strip are evaluated and are
converted into voltage signals using the transfer function associated to the detector electronics, taking into account the
detector noise as well as the noise associated to the electronics. The
fired strips and the time over threshold are then determined.

The signals in the Tracker are then analysed by the reconstruction package. It generates a series of clusters, that are used to find and fit the best track candidates.
This last procedure is done using alternative pattern recognition algorithms and a Kalman Filter based algorithm. Finally, using the best track found, another algorithm finds the best vertex candidate for gamma events.

\begin{figure}[ht!]
\centering
\includegraphics[width=3.in]{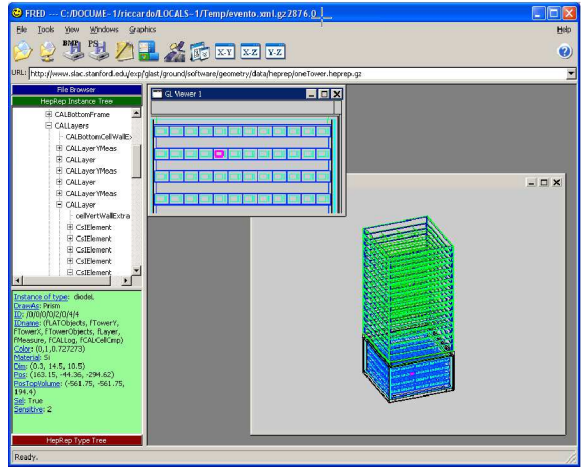}
\caption{GLAST LAT event display based on FRED}
\label{fig--fred}
\end{figure}

Although it is not part of the simulation, the visualization package is essential for the use of the simulation itself.
A new version~\protect\cite{Frailis:2003yp,fredweb} of the event display based on the HepRep protocol was developed and integrated in the offline software.
The figure~\ref{fig--fred} shows a recent FRED-based event display of GLAST. 

\section{Simulating the GLAST LAT High Energy Sky}

\begin{figure}[ht!]
\centering
\includegraphics[width=4.in]{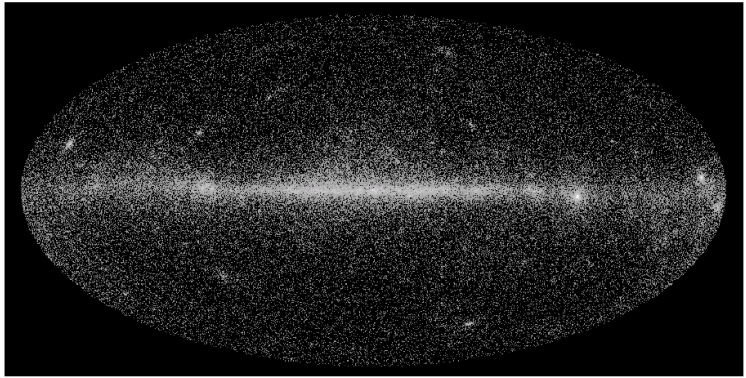}
\caption{Full simulated gamma-ray sky seen by GLAST LAT}
\label{lat-dc1}
\end{figure}

Using Gleam it's possible to compute and store time, direction and energy of each incoming simulated and reconstructed gamma ray. The Data Challenge One (DC1), organized by the LAT Collaboration from December 2003 to February 2004, represented the first opportunity to test the complete simulation chain, and the first attempt to perform scientific analyses on simulated data. 
For DC1, only the gamma-ray sky was simulated, while the cosmic-ray flux (about $10^4$ times greater) was modeled separately for development of the background rejection algorithms. These algorithms were then applied to the simulated gamma-ray data.
The gamma-ray sky simulated for DC1 was quite rich: a variety of sources was included. The simulation software takes into account the relative fluxes and computes from which source the next photon arrives.
A map of the simulated sky is reported in fig.~\ref{lat-dc1}. 

Based on the simulated data, several groups tried to develop the best algorithms to detect unknown gamma-ray sources. GLAST observations contain photons from astrophysical sources, convoluted with the spatial and spectral instrument response. Moreover in most astronomical gamma-ray images a large fraction of sources is near the detection limit. Thus a careful statistical treatment is needed to determine their existence and properties. Many tools (parametric methods) need a priori model to fit the data and estimate their parameters. No model or hypotesis on the data are requested by the wavelet method\protect\cite{marcucci}. Through an iterative procedure the wavelet method allows a blind search of gamma-ray point~sources.
 
Another goal for the DC1 was to test the possibility to trigger Gamma-Ray Bursts using only LAT information. 
Several groups within the LAT collaboration are prototyping trigger and alert algorithms for detecting transient signals in DC1 data, and different algorithms were studied\protect\cite{longo05}. 
Different algorithms were successfully applied for searching for transient signals in DC1 data. Bright bursts (with fluence greater than $10^{-5}$ erg/cm$^2$ between 20 keV and 1 MeV) can be detected with simple algorithms. Further studies will include the particle background, and the possibility to implement an on-board LAT alert algorithm. All of these items will be addressed for the next Data Challenge (DC2), in which one month of simulated data will be produced.

\section{Conclusions}

The {\it Gleam} simulation program has been developed in the last few years and now it's ready for simulating the full GLAST satellite and is being used for deriving the final instrumental parameters  and for generating a full set of events for the developing of scientific analysis software.

\end{document}